\documentclass[]{aa}
\usepackage{graphicx,amsmath,times,natbib,dcolumn,xspace}

\bibpunct{(}{)}{;}{a}{}{,} 
\def\5{\footnotesize V\normalsize}
\def\4{\footnotesize IV\normalsize}
\def\3{\footnotesize III\normalsize}
\def\2{\footnotesize II\normalsize}
\def\1{\footnotesize I\normalsize}
\def\lam{$\lambda$}
\def\lamlam{$\lambda\lambda$}
\def\kms{$\mbox{km~s}^{-1}$\xspace}

\def\v{$\phantom{^{l}}$}

\def\lsim{\mathrel{\rlap{\lower4pt\hbox{\hskip1pt$\sim$}}
    \raise1pt\hbox{$<$}}}    
\def\gsim{\mathrel{\rlap{\lower4pt\hbox{\hskip1pt$\sim$}}
    \raise1pt\hbox{$>$}}} 

\begin{document}

\title{The VLT-FLAMES survey of massive stars: NGC346-013 as a test case for massive close binary evolution\thanks{Based on observations at the European Southern Observatory, Paranal, Chile 
in programmes 171.D-0237 and 081.D-0364}}

\author{
B.~W.~Ritchie\inst{1,2},
V.~E.~Stroud\inst{1,3,4},
C.~J.~Evans\inst{5},
J.~S.~Clark\inst{1},
I.~Hunter\inst{6}, 
D.~J.~Lennon\inst{7}, 
N.~Langer\inst{8,9}, 
and S.~J.~Smartt\inst{6}
}

\offprints{b.ritchie@open.ac.uk}
\authorrunning{B.~W.~Ritchie~et~al.}
\titlerunning{VLT-FLAMES survey of massive stars: NGC346-013}
   
\institute{       
 Department of Physics and Astronomy, The Open University, Walton
 Hall, Milton Keynes MK7 6AA, UK
\and
 Lockheed Martin Integrated Systems, Building 1500, Langstone, Hampshire, PO9 1SA, UK
\and
 Faulkes Telescope Project, School of Physics and Astronomy, Cardiff University, Cardiff, CF24 3AA, UK
\and
 Division of Earth, Space and Environment, University of Glamorgan, Pontypridd, CF37 1DL, UK
\and
 UK Astronomy Technology Centre, Royal Observatory Edinburgh, Blackford Hill, Edinburgh, EH9 3HJ, UK
\and
 Department of Physics \& Astronomy, Queen's University Belfast, Belfast BT7 1NN
, Northern Ireland, UK
\and
 ESA/STScI, 3700 San Martin Drive, Baltimore, MD 21218, USA
\and       
 Argelander-Institut f\"ur Astronomie der Universit\"at Bonn, Auf dem H\"ugel 71, 53121 Bonn, Germany
\and 
 Astronomical Institute, Utrecht University, Princetonplein 5, Utrecht, The Netherlands
}
\date{Received 16 July 2011 / Accepted 20 October 2011 }

\abstract
  {NGC346-013 is a peculiar double-lined eclipsing binary in the Small
    Magellanic Cloud (SMC) discovered by the VLT-FLAMES survey of
    massive stars.}
  {We use spectroscopic and photometric observations to investigate
    the physical properties and evolutionary history of NGC346-013.}
  {Spectra obtained with VLT/FLAMES are used to construct a radial
    velocity curve for NGC346-013 and to characterise the early B-type
    secondary. Photometry obtained with the Faulkes Telescope South is
    then used to derive orbital parameters, while spectra of the
    secondary are compared with synthetic spectra from TLUSTY model
    atmospheres.}
  {The orbital period is found to be 4.20381(12) days, with masses of
  $19.1\pm1.0$ and $11.9\pm0.6M_\odot$.  The primary is a rapidly
  rotating ($v_\text{rot}=320\pm30$\kms) late-O dwarf while the
  secondary, an early-B giant, displays synchronous rotation and has
  filled its Roche lobe, implying that it was originally the more
  massive component with recent mass transfer `spinning up' the
  primary to near-critical rotation. Comparison with synthetic spectra
  finds temperatures of 34.5kK and 24.5kK for the primary and
  secondary respectively, with the nitrogen abundance of the
  secondary enhanced compared to baseline values for the SMC,
  consistent with the predictions of models of interacting binaries.}
{NGC346-013 likely evolved via non-conservative mass transfer in a
  system with initial masses $\sim$22+15$M_\odot$, with the
  well-constrained orbital solution and atmospheric parameters making
  it an excellent candidate for tailored modelling with binary
  evolution codes. This system will form a cornerstone in constraining
  the physics of thermal timescale mass transfer, and the associated
  mass transfer efficiency, in massive close binary systems.}


\maketitle 

\section{Introduction}\label{intro}

Binary systems are of critical importance for understanding the
formation and evolution of massive stars. They permit both the direct
determination of fundamental stellar properties such as mass and
radius, and, through properties such as orbital period, eccentricity
and mass ratio, offer insight into the processes of massive star
formation and dynamical interactions in the natal cluster. Binarity
will strongly influence the evolutionary path both components will
follow, with tidal interaction affecting rotational mixing
\citep{demink} and transfer of mass and angular momentum in the later
evolutionary stages leading to a distribution of spectral types that
differs significantly from single-star population models
\citep{eldridge}. Ejection of the primary's hydrogen mantle during
close binary evolution will shape both the type of supernova and the
nature of the subsequent relativistic object \citep{wellstein, brown},
while massive secondaries spun-up to critical rotation during mass
transfer represent candidates for $\gamma$-ray burst progenitors
\citep{woosley,cantiello}. Properties derived from massive binary
systems may also provide key constraints on the formation and
evolution of wider stellar populations \citep{ritchie10, clark11}.

In this paper, we examine the short-period eclipsing binary
\object{NGC346-013}\footnote{RA=00:59:30.27,
  $\delta$=$-$72:09:09.3~(J2000); \object{star~\#782} from \cite{mpg},
  who list $m_v=14.46$, $(B-V)=-0.18$, and a reddening-free
    colour index $Q=(U-B)-0.72(B-V)=-0.79$} discovered in the
VLT-FLAMES survey of massive stars\footnote{\cite{evans05,evans06}.}
field centred on \object{NGC\,346}, a young, massive cluster located
at the centre of \object{N66} \citep{henize56}, the largest H~\2
region in the Small Magellanic Cloud (SMC). The region has undergone
extensive star formation in recent times, with \object{NGC\,346}
hosting the largest population of O-type stars in the SMC \citep{mpg,
  wal00,evans06}, while a large population of massive young stellar
objects (YSOs) indicate that star formation is ongoing \citep{simon07,
  g10}. Studies of pre-main sequence stars indicate a complex history,
with major star formation in \object{NGC\,346} starting $\sim$6Myr ago
and lasting about 3Myr, while an older population indicates an earlier
epoch of star formation that took place $\sim$10Myr ago \citep{h08,
  cig10a, cig10b}. 

The main morphological features seen in the FLAMES spectra of
\object{NGC346-013} were summarized by \cite{evans06}, who noted lines
of \ion{He}{i}, \ion{C}{iii}, \ion{N}{iii} and \ion{Si}{iii}
consistent with a spectral type of B1--1.5 and velocity shifts between
spectra that spanned $\sim$400\kms. However, a broad
\ion{He}{ii}~\lam4686 absorption line was observed moving in the
opposite sense to the other features, indicating the presence of a
companion with an earlier spectral type, with the lower amplitude of
the line indicating that it traces the more massive object in the
system: throughout this paper we refer to the hotter, more massive
object as the primary and the B-type star as the secondary,
acknowledging the problems in applying these terms to evolved,
short-period binary systems in which mass exchange may confuse their
conventional meaning.

The peculiar hot, underluminous primary and the presence of an
apparently rotationally-broadened He~II line suggest that
\object{NGC346-013} is seen in a brief evolutionary phase shortly
after the end of thermal-timescale Roche-lobe overflow, where mass
transfer has reversed the initial mass ratio of the system and
`spun-up' the mass-gainer to rapid rotation. Such a system is of
critical importance to understanding the efficiency of mass transfer
in short-period binaries \citep{wellstein,wellstein01,p05,demink07}, a
process that shapes both the ongoing evolution of the systems and the
extreme evolutionary endpoints that can only be reached via extensive
binary interaction, such as \object{Wray~977} (a $\sim$40+1.4$M_\odot$
B1~Ia$^+$/neutron star binary; \citealt{wellstein}; \citealt{kaper}),
\mbox{\object{4U1700-37}} (a 58+2.4$M_\odot$ O6.5~Iaf$^+$/compact
object binary; \citealt{clark02}) or the magnetar
\object{CXOU~J164710.2-455216}, which formed a neutron star despite a
minimum progenitor mass approaching 50$M_\odot$ \citep{ritchie10}.  We
have therefore undertaken a programme of observations aimed at
deriving accurate properties for the two components of
\object{NGC346-013}, supplementing the original dataset discussed by
\cite{evans06} with an additional 14 spectra obtained with FLAMES in
2008 and a photometric dataset acquired during 2008--2009 with the
Faulkes Telescope South (FTS).  Details of our observations are given
in Sect.~\ref{sec:obsinfo}, while in Sect.~\ref{sec:tlusty} we use
{\sc tlusty} model atmospheres \citep[e.g.][]{hl95} to estimate the
temperature, gravity, rotational velocity and chemical abundances of
the early-B secondary that dominates the FLAMES spectra. In
Sect.~\ref{sec:orbit} we derive orbital parameters from the radial
velocity (RV) and photometric datasets, and discuss the evolution of
\object{NGC346-013} and implications for \object{NGC\,346} in
Sect.~\ref{sec:discuss}.

\section{Observational Details}
\label{sec:obsinfo}

\begin{table}
\caption{FLAMES setups and lines used for RV measurement.}
\label{tab:flames}
\begin{center}
\begin{tabular}{lccl}
Setup & Coverage ($\text{\AA}$) & $R (\lambda_c)$  & Lines \\
\hline\hline
HR02   & 3854--4049 & 19600 & \ion{He}{i}~\lam4009, \ion{He}{i}~\lam4026\\ 
HR03   & 4033--4201 & 24800 & \ion{He}{i}~\lam4121, \ion{He}{i}~\lam4144\\
HR04   & 4188--4392 & 20350 & --\\
HR05   & 4340--4587 & 18470 & \ion{He}{i}~\lam4388, \ion{He}{i}~\lam4471\\
       &            &       & \ion{Si}{iii}~\lamlam4553, 4568\\
HR06   & 4538--4759 & 20350 & \ion{Si}{iii}~\lamlam4553, 4568\\
       &            &       & \ion{N}{iii}~\lam4642, \ion{C}{iii}~\lam4650\\
       &            &       & \ion{He}{ii}~\lam4686, \ion{He}{i}~\lam4713\\
\hline
\end{tabular}
\end{center}
\end{table}

\begin{table}
\caption{Journal of observations.}
\label{tab:observations}
\begin{center}
\begin{tabular}{ccc D{,}{\,\pm\,}{-1} D{,}{\,\pm\,}{-1} }
            &           & GIRAFFE & \multicolumn{1}{c}{RV$_\text{sec}$} &  \multicolumn{1}{c}{RV$_\text{pri}$} \\  
MJD$^a$     & Phase$^b$ & Setup   & \multicolumn{1}{c}{(\kms)} & \multicolumn{1}{c}{(\kms)} \\  
\hline\hline
52926.08623 & 0.445     & HR06    & 245,7  &  \\ 
52926.13820 & 0.457     & HR06    & 244,20 & 152,16 \\ 
52926.16522 & 0.464     & HR06    & 226,10 & 144,15 \\ 
52926.20344 & 0.473     & HR06    & 211,13 & 111,17 \\ 
52926.23039 & 0.479     & HR06    & 201,5  & 144,16 \\ 
52954.13589 & 0.117     & HR02    & 313,7  & 47,6^c \\ 
52954.16291 & 0.124     & HR02    & 335,9  & \\ 
52954.18986 & 0.130     & HR02    & 344,10 & 47,7^c\\ 
52955.08957 & 0.344     & HR02    & 370,9  & \\ 
52955.11657 & 0.351     & HR02    & 348,27 & 40,9^c\\ 
52955.14352 & 0.357     & HR02    & 337,14 & \\ 
52978.11549 & 0.822     & HR05    & -59,5  & 290,7^c\\ 
52978.14244 & 0.828     & HR05    & -59,12 & 275,7^c\\ 
52978.16940 & 0.835     & HR05    & -46,14 & 266,6^c\\ 
52981.04545 & 0.519     & HR03    & 139,5  & \\ 
52981.07248 & 0.525     & HR03    & 116,8  & \\ 
52981.09943 & 0.532     & HR03    & 96,5   & \\ 
52981.13275 & 0.539     & HR03    & 91,10  & \\ 
52981.15970 & 0.546     & HR03    & 69,11  & \\ 
52981.18665 & 0.552     & HR03    & 70,5   & \\ 
52988.11104 & 0.199     & HR06    & 396,17 & -10,11\\ 
52988.15506 & 0.210     & HR06    & 412,13 & -10,13\\ 
52988.18332 & 0.217     & HR06    & 406,15 & -15,12\\ 
53006.04733 & 0.466     & HR05    & 226,5  & \\ 
53006.07434 & 0.473     & HR05    & 194,8  & \\ 
53006.10131 & 0.479     & HR05    & 202,15 & \\ 
\hline
54663.31918 & 0.693     & HR02    & -76,5  & 293,9^c\\
54663.35111 & 0.701     & HR02    & -78,6  & 304,11^c\\
54664.29465 & 0.925     & HR02    & 64,5   & \\
54664.32772 & 0.933     & HR02    & 62,5   & \\
54665.29193 & 0.163     & HR02    & 378,5  & -2,7^c \\
54666.25561 & 0.392     & HR02    & 326,6  & \\
54666.28778 & 0.400     & HR02    & 294,5  & \\
54666.35881 & 0.416     & HR02    & 286,5  & \\
54666.39125 & 0.424     & HR02    & 266,5  & \\
54669.36147 & 0.131     & HR02    & 342,13 & 28,8^c \\
54669.39421 & 0.138     & HR02    & 356,10 & \\
54670.28291 & 0.350     & HR02    & 391,9  & \\
54671.26508 & 0.584     & HR02    & 42,8   & \\
54672.28464 & 0.826     & HR02    & -66,9  & 294,9^c\\
\hline
\end{tabular}
\end{center}
$^a$MJD given at the midpoint of the integration, $^b$For superior
conjunction at MJD 52924.21553(66) and orbital period 4.20381(12) days
(see Sect.~\ref{sec:radial}), $^c$From two-component Gaussian
  fits to the \ion{He}{i} lines.
\end{table}

\subsection{Spectroscopy}

A total of 32 spectra of \object{NGC346-013} were obtained in
2006 as part of the VLT-FLAMES survey of massive stars field centred
on \object{NGC~346} \citep{evans06}. Observations were taken with the
GIRAFFE spectrograph, with the five relevant wavelength settings used
listed in Table~\ref{tab:flames}\footnote{A further setting,
    HR14A (covering 6391--6701$\text{\AA}$) is described by
    \cite{evans06} but is not used here, as H$\alpha$ and
    \ion{He}{i}~\lam6678 are strongly blended with nebular
    emission. Observations in HR04 mode were not used for RV
    measurement due to a lack of strong metal lines in this region of
    the spectrum, but the H$\delta$ line was used for surface gravity
    fits, as discussed in Sec.~\ref{atmos}.}. An integration time of
2275s was used, with the exception of HR06~\#08 (MJD=52988.18332)
which used an integration time of 2500s (related to execution of the
service mode observations).  The data were reduced using version 1.10
of the Giraffe Base-Line Reduction Software (girBLDRS;
\citealt{blecha}), with sky spectra subtracted from each target using
the \textit{Starlink} package {\sc DIPSO} \citep{how03}. A further 14
spectra in HR02 mode were obtained during a 9-day period in July~2008
as part of a programme of follow-up observations of the cluster
(PI:Hunter), each with an integration time of 2662s. These data were
pipeline reduced using version 3.6.8 of {\sc
  esorex}\footnote{http://www.eso.org/sci/data-processing/software/cpl/esorex.html},
with individual spectra shifted to the heliocentric frame using the
IRAF\footnote{IRAF is distributed by the National Optical Astronomy
  Observatories, which are operated by the Association of Universities
  for Research in Astronomy, Inc., under cooperative agreement with
  the National Science Foundation.} {\sc rvcorrect} and {\sc dopcor}
routines. A median sky spectrum was then created from the sky fibres
and subtracted from each object spectrum. The signal-to-noise
  ratio was in excess of $\sim$75, varying slightly depending on the
  exact conditions when individual observations were obtained.

\begin{figure}
\begin{center}
\resizebox{\hsize}{!}{\includegraphics{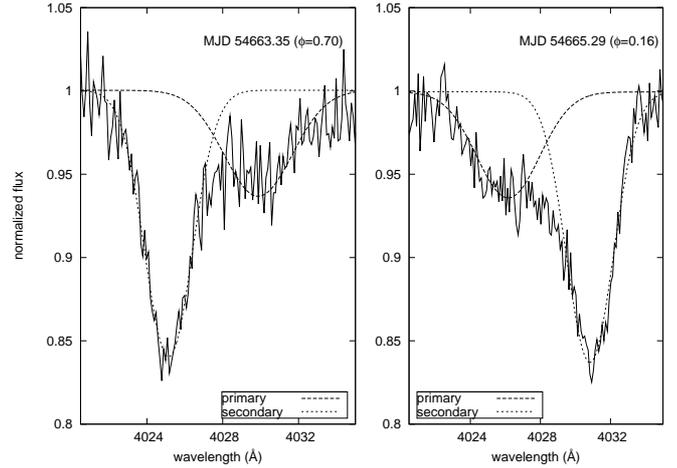}}
\caption{Two-component Gaussian fit to the
  \ion{He}{i}~\lam4026 lines near maximum separation at MJD54663.35
  ($\phi=0.70$; {\it left panel}) and approximately half an orbit
  later at MJD54665.29 ($\phi=0.16$; {\it right
    panel}). }\label{fig:deconv}
\end{center}
\end{figure}

The strong Balmer-series lines could not be used for RV measurement
due to their broad wings, lack of sharp core, and presence of
nebular emission, and RVs were instead measured using the strongest
available helium and (when present) metal absorption lines in each
spectrum: the lines used for each setup are listed in
Table~\ref{tab:flames}. The \ion{He}{i} lines appear to show only
minimal nebular components, but are notably asymmetric in most spectra
with a strong, narrow and symmetric core arising in the secondary and
a weaker contribution from the hotter primary in the wings;
example profiles near maximum separation are shown in
Fig.~\ref{fig:deconv}.  RVs for the secondary were therefore measured
by fitting profiles to just the core of the strong \ion{He}{i}
component, with the derived RVs in excellent agreement with other
lines of \ion{C}{iii} and \ion{Si}{iii} that appear free from strong
contamination by the primary. The derived RV for each spectrum is then
an error-weighted average of individual absorption lines. Unblended
lines from the primary are only directly visible in the HR06 data,
with a comparison of spectra from 2003 October 14 and 2003 December 15
(MJDs of 52926 and 52988 respectively) clearly showing the antiphase
motion of the \ion{He}{ii}~\lam4686 line (see Fig.~4 of
\citealt{evans06}), with the weakness of the corresponding
He~II~\lam4542 absorption line suggests a classification around
O9--9.5. However, the HR06 spectra sample only a small portion of the
orbit ($\phi\sim$0--0.25), and we therefore use two-component Gaussian
fits to the the strong \ion{He}{i}~\lam4026 and \ion{He}{i}~\lam4471
lines in the HR02 and HR05 spectra taken near maximum separation to
further constrain the amplitude of the primary. The \ion{He}{i}
profiles are clearly double at these epochs, and satisfactory fits to
co-added spectra can be achieved\footnote{However, in a number of
  spectra sky residuals fall within the weak primary component of the
  \ion{He}{i} line, precluding accurate deconvolution despite the
  primary and secondary being well separated; these spectra are not
  used in the analysis.}. A full list of derived RVs for both primary
and secondary are listed in Table~\ref{tab:observations}.

\subsection{Photometry}
\label{sec:photo}

Photometry was obtained with the Faulkes Telescope South (FTS) 2m
Ritchey-Chr\'{e}tien Cassegrain, located in Siding Spring, Australia,
as part of a follow-up photometric programme to monitor the target
clusters of the FLAMES survey. FTS currently uses a Merope camera
(EM03), although prior to 2009 January the camera used was an Apogee
`Hawkcam' (EA02). Both cameras were coupled with an EV2 CCD42-40DD CCD
giving a $4.7'\times4.7'$ field of view and producing images of
$2048\times2048$ pixels, binned $2\times2$ to give $1024\times1024$
pixels at $0.278''$~pixel$^{-1}$.  The source was observed using a
Bessel $V$ filter, with 100s exposures. Science images were produced
using the Faulkes automatic pipeline, which de-biases and flat-fields
the raw images.

A total of 104 observations of \object{NGC\,346} were obtained through
the offline queue between 2008 January 10 and 2009 December 23, with
seeing values ranging from 0\farcs4 to 3\farcs1.  Aperture
differential photometry of \object{NGC346-013} and two nearby cluster
members, \object{MPG~781} (=\object{Sk~80}, \object{NGC346-001}) and
\object{MPG~811}, was performed using IRAF package {\sc apphot}. Two
moderately bright stars lie less than $3''$ to the south-east of
\object{NGC346-013} and, at the pixel scale of the FTS data, the
three sources could not be separated by using aperture photometry
alone. To overcome this a large 12-pixel aperture was used to include
all three sources, making the assumption that the other two stars are
non-variable and therefore any variability is solely due to the
binary. The relative contribution of the target with respect to the
other two stars was then measured by obtaining point spread function
(PSF) photometry of one of the observations under excellent seeing,
using {\sc iraf/daophot}.  The two stars were found to contribute
$34\%$ of the light at phase $\phi=0.7$.

\section{Quantitative spectroscopic analysis of system}
\label{sec:tlusty}

\begin{figure}
\centering
\begin{tabular}{c}
\includegraphics[height=60mm,angle=0]{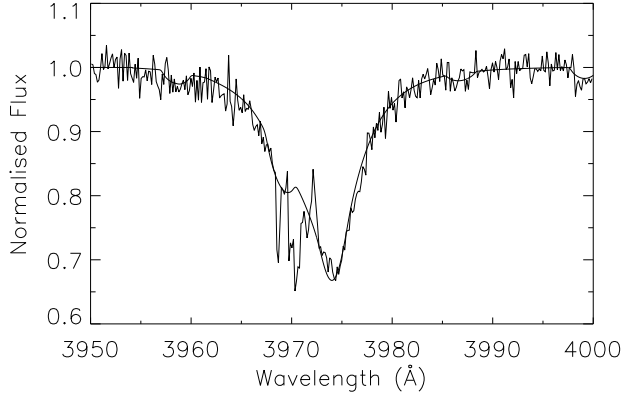}\\
\includegraphics[height=60mm,angle=0]{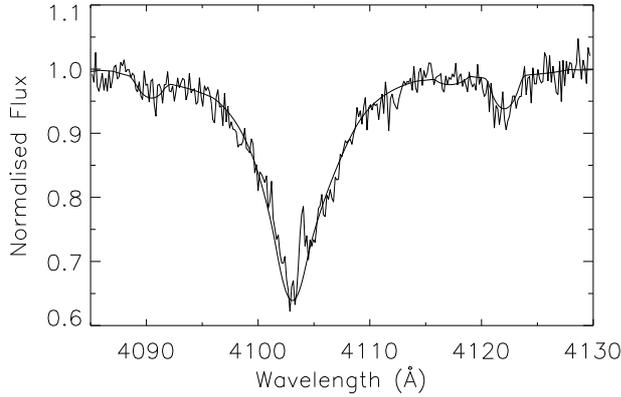}\\
\includegraphics[height=60mm,angle=0]{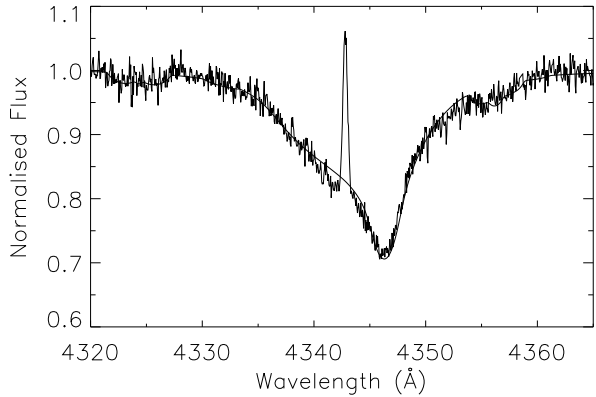}\\
\end{tabular}
\caption[]{Quality of model fits to the observed H$\epsilon$, H$\delta$ and H$\gamma$ lines 
(upper, middle and lower panels respectively) of both stars, with 
the B-star being 1.8 times brighter than the O-star. The
absorption features in the blue flank of the H$\epsilon$ line are due
to interstellar Ca~II~$\lambda$3968.}
\label{hlines}
\end{figure}

\begin{figure}
\centering
\includegraphics[height=60mm,angle=0]{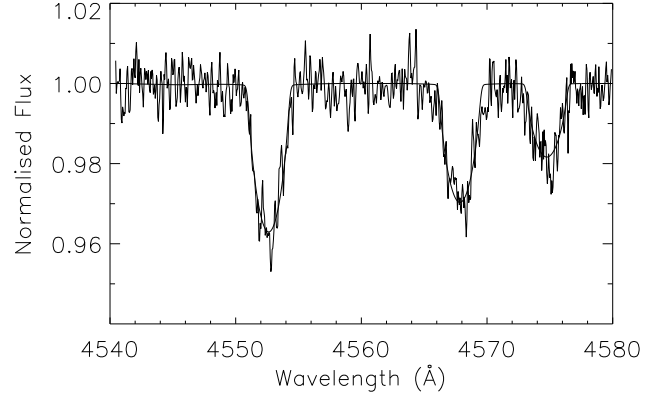}\\
\includegraphics[height=60mm,angle=0]{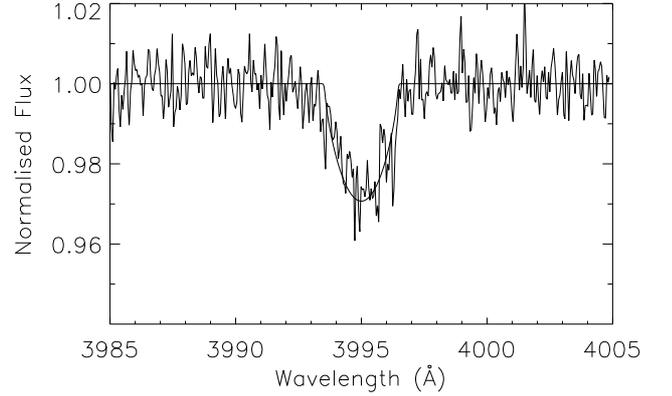}\\
\caption[]{Example fits to the metal absorption lines: (Top) \ion{Si}{iii} triplet (bottom) \ion{N}{ii} 3995 line.}
\label{f_fits}
\end{figure}

\subsection{Atmospheric parameters}\label{atmos}

To determine effective temperatures ($T_\text{eff}$) and surface
gravities ($\log g$) for the orbital analysis, physical parameters for
the secondary object were estimated using the non-LTE {\sc tlusty}
model atmosphere grid, discussed by \citet{duf05} and used extensively
in the analysis of B-type Magellanic Cloud stars from the FLAMES
survey (e.g. \citealt{ih06, t07}).  Standard techniques were used to
derive the physical parameters of the secondary. The effective
temperature was estimated from the \ion{Si}{iii}/\ion{Si}{iv}
ionization equilibrium, and fits to the hydrogen lines were used to
constrain the surface gravity. The \ion{Si}{iii} triplet at
4560\AA\ was used to estimate the microturbulence (see \citealt{ih06}
for a detailed discussion). Deriving the atmospheric parameters is an
iterative process and is further complicated by contamination of the
spectra of the secondary by the O-type primary, and an estimate of the
continuum contamination must therefore be included. The fitting of the
hydrogen lines for the gravity estimate is dependent on the level of
continuum contamination from the primary object, and a secondary 1.8
times brighter than the primary provides the best fit to the observed
spectra (i.e. a continuum contamination of 35\%). As the spectra of
the primary and secondary objects are well separated in the HR02 and
HR04 observations, the H$\epsilon$ and H$\gamma$ lines have been used
to determine the surface gravity of the secondary object, with the best
estimates being 3.40\,dex and 3.50\,dex from the two lines
respectively. The individual fits to both these lines as well as the
H$\delta$ line where the primary and secondary are not well separated
are shown in Fig.~\ref{hlines} and provide satisfactory fits in all
cases. We derive T$_{\rm eff}$\,$=$\, 24\,500\,K in reasonable
agreement with the temperature scale from \cite{t07} for early B-type
giants in the SMC.  A microturbulence ($\xi$) of 4\,km\,s$^{-1}$ was
estimated from the \ion{Si}{iii} triplet of lines.  Conservative
uncertainties on these results are $\Delta T_{\rm
  eff}$\,$=$\,$\pm$1500\,K, $\Delta \log g$\,$=$\,$\pm$0.2, and
$\Delta \xi$\,$=$\,$\pm$3\,km\,s$^{-1}$.

Simultaneously to the analysis of the secondary object, the
spectroscopic analysis of the primary object was carried out. The
rotational velocity of the primary is much greater than the secondary
(Sec.~\ref{sec:rotvel}) which precludes the observation of the metal
lines in its spectra. As such, the \ion{He}{ii} lines were utilised
for the effective temperature estimate of the primary object by
fitting the observed line with rotationally broadened model spectra
over a range of parameters, with a best fit to the \ion{He}{ii}
4542\AA\ line found at an effective temperature of 34\,500\,K.  The
logarithmic gravity of the primary was calculated to be 3.90\,dex from
the fit to the hydrogen lines. Due to the weakness of the primary
spectrum, errors are $\Delta T_{\rm eff}$\,$=$\,$\pm$3000\,K and
$\Delta \log g$\,$=$\,$\pm$0.3 respectively.

\subsection{Rotational velocities}\label{sec:rotvel}

Similar methods to those discussed in \cite{ih06} have again been used
to estimate the rotational velocity of both the primary and secondary
objects, viz. convolving the model spectra with rotationally-broadened
profiles until a good fit to the observations is found.  No metal
lines are apparent from the primary, so we have used the \ion{He}{ii}
line at 4686\AA\ to estimate a projected rotational velocity of
320~\kms for this object\footnote{This value is in excellent agreement
  with the fits to the \ion{He}{i} lines used to measure the RV of the
  primary.}. A model profile at the estimated parameters discussed
above was scaled to the same equivalent width as the observations in
order to derive this value. However as the profile is rotationally
dominated, errors in the underlying model spectrum should not be
significant; nevertheless we adopt a conservative uncertainty of
30~\kms. Assuming a continuum contamination of 35\% from the primary
we derive a projected rotational velocity of 110$\pm$10~\kms for the
secondary object from the \ion{Si}{iii} triplet of lines.

\subsection{Chemical Abundances}\label{sec:abund}

\begin{table}
\begin{center}
\caption[]{Equivalent widths for the available 
metallic lines of the secondary.}
\label{t_lineabund}
\begin{tabular}{lcccc}
Species&$\lambda$ (\AA)& EW$_{\rm meas.}$ (m\AA)   &EW$_{\rm corr.}$ (m\AA)& [$X$/H]  \\
\hline
\hline
N II  & 3995  &62 & \v98  & 7.54\\ 
O II  & 4069  &89 & 129   & 8.19\\
O II  & 4072  &65 &  \v99  & 8.34\\
O II  & 4076  &75 &  117  & 8.35\\
O II  & 4591  &51 & \v80  & 8.21\\
O II  & 4596  &48 & \v75  & 8.23\\
O II  & 4662  &42 & \v66  & 8.04\\
Mg II & 4481&53 & \v34  & 6.68\\
Si III& 4553  &87&  142  & 6.88\\
Si III& 4568  &69&   108 & 6.85\\
Si III& 4575  &42&  \v66  & 6.90\\
Si IV & 4116 &25& \v39  & 6.85\\
\hline
\end{tabular}
\end{center}
{\it Note:} Abundances (where [$X$/H] = log($X$/H)$+$12) have been
calculated by scaling the measured equivalent widths (EW$_{\rm meas.}$)
to give corrected widths (EW$_{\rm corr.}$) assuming that the secondary 
is 1.8 times brighter than the primary.
\end{table}

\begin{table}
\begin{center}
\caption{Atmospheric parameters for the secondary of
  NGC\,346-013}\label{t_avabund}
\begin{tabular}{llc}
        & NGC346-013 & SMC average\\
\hline
\hline
T$_{\rm eff}$ (K)    & $24\,500 \pm 1\,500$ & $-$    \\
$\log g$             & $3.45\pm0.2$      & $-$    \\
$v_{\rm turb}$ (\kms) & $4\pm3$          & $-$    \\
${\rm [N/H]}$       & $7.54\pm0.18$    & 6.56\\
${\rm [O/H]}$       & $8.23\pm0.23$    & 8.02\\
${\rm [Mg/H]}$      & $6.68\pm0.16$    & 6.72\\
${\rm [Si/H]}$      & $6.87\pm0.30$    & 6.77\\
\hline
\end{tabular}
\end{center}
{\it Note}: Mean abundances (log($X$/H)$+$12) and the uncertainties in
these parameters are given. For comparison the mean SMC abundances
from \cite{ih06} are also included.
\end{table}

Using the above parameters, chemical abundances for the secondary
object have been determined (N, O, Mg and Si). The equivalent widths
of the observed metal lines of the secondary have been measured and
these measured equivalent widths are scaled to take into consideration
the continuum contamination from the primary and then the abundances
are estimated. The line-by-line equivalent widths and abundance
estimates are given in Table~\ref{t_lineabund} with the mean values
being summarized in Table~\ref{t_avabund}. Additionally we give the
base-line chemical composition of the SMC as determined by \cite{ih06}
from the analysis of 14 early B-type objects. The O, Mg and Si
abundances are in good agreement which further supports our derived
continuum contamination parameter. Although the nitrogen abundance is
$\sim$1.0\,dex enhanced over that of the SMC, \cite{ih06,ih09}
determined nitrogen abundances ranging from 6.5\,dex to 7.6\,dex and
so the enhanced nitrogen abundance of the secondary in
\object{NGC346-013} is not exceptional. In Fig.~\ref{f_fits} we
show model fits to example metallic lines.

The uncertainties given in Table~\ref{t_avabund} for the abundances
include random uncertainties arising from measurement errors as well
as the systematic uncertainties arising from errors in the atmospheric
parameters, see \cite{ih06} for a full discussion on the derivation of
these uncertainties.

\section{Orbital Parameters}
\label{sec:orbit}

\begin{figure}
\begin{center}
\resizebox{\hsize}{!}{\includegraphics{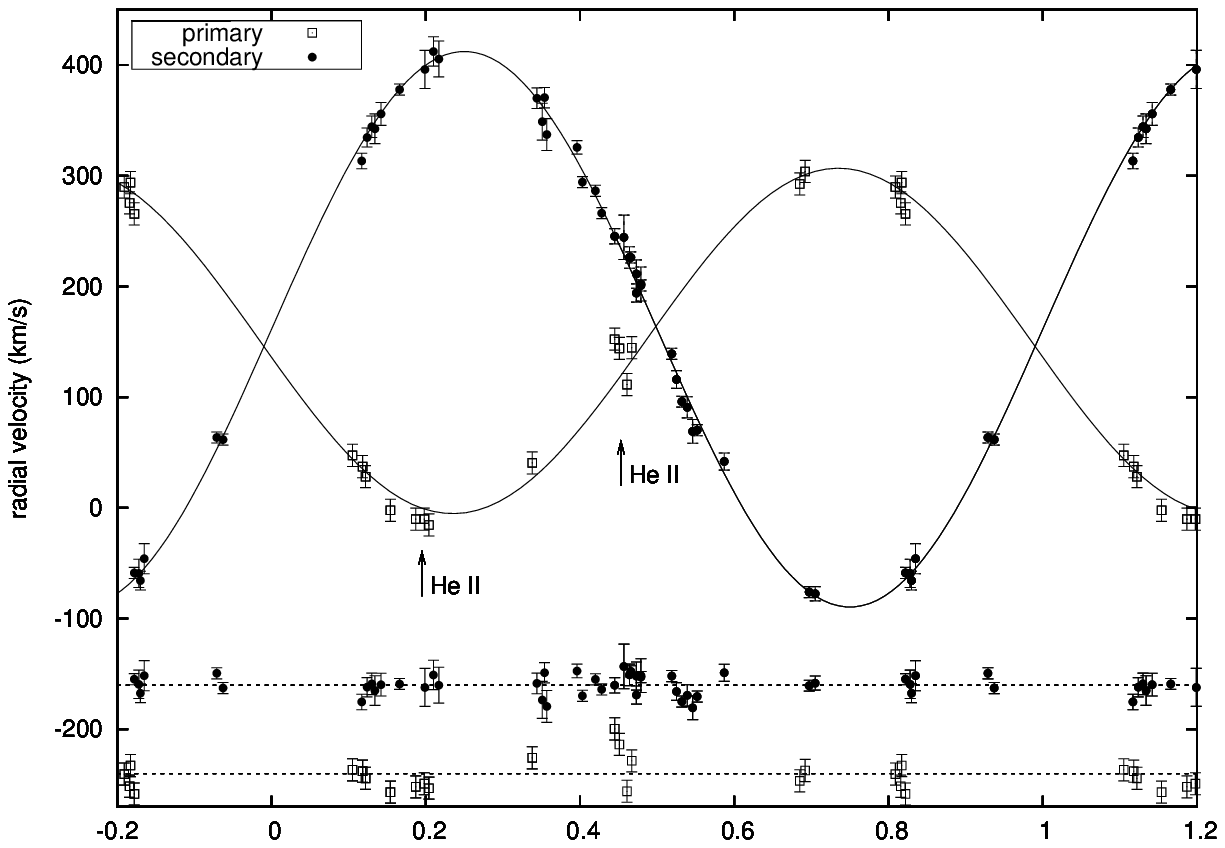}}
\resizebox{\hsize}{!}{\includegraphics{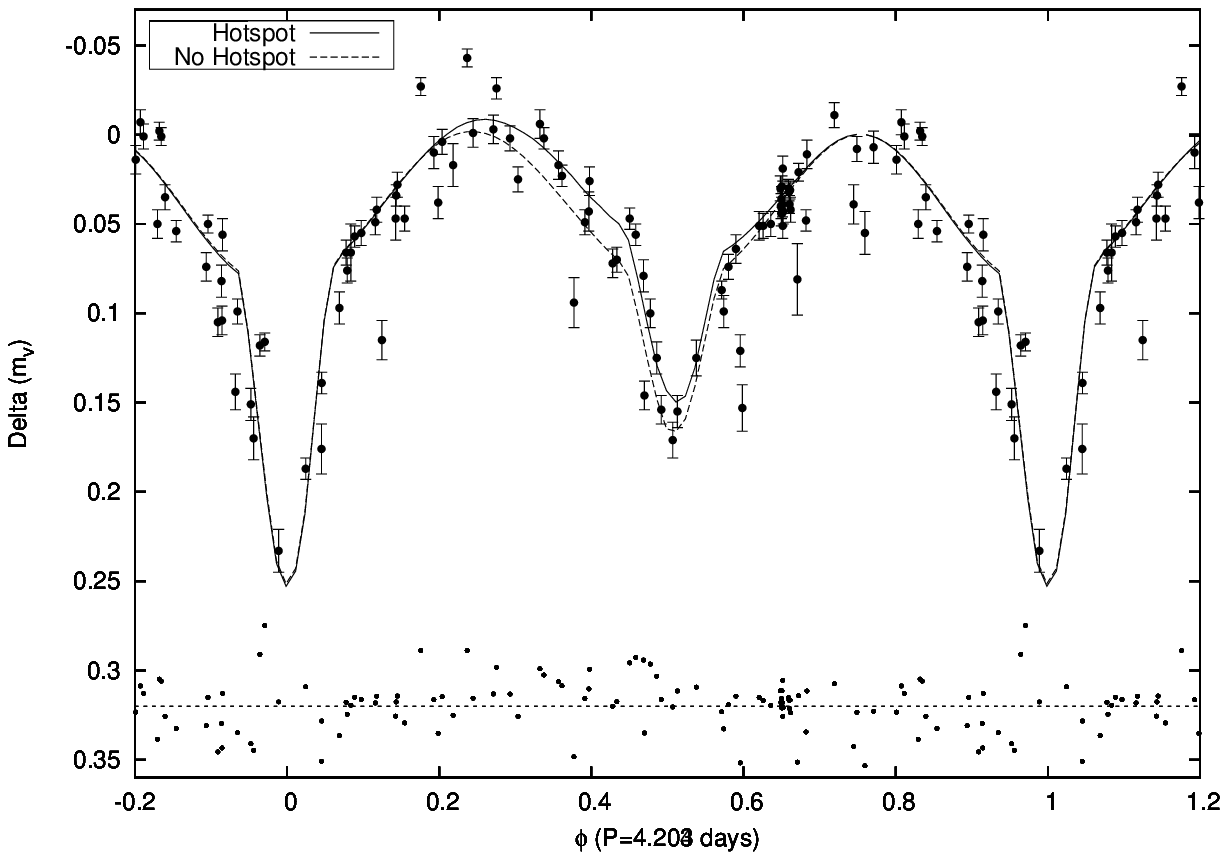}}
\caption{(Top) Radial velocity curve for \object{NGC346-013}.
(bottom) \textit{Nightfall} fits to the $V$-band lightcurve. The
solid and dashed lines show models with and without the hotspot
discussed in the text; residuals are plotted for the no-hotspot model.}
\label{fig:phase_lightcurve}
\end{center}
\end{figure}

\subsection{Radial velocity curve}
\label{sec:radial}

An initial estimate of the orbital period was obtained using separate
Lomb-Scargle periodograms of the radial velocity and photometric data,
with the strongest peak in both datasets found at 0.238 cycles/day
(corresponding to an orbital period of 4.202 days), with the 2008
FLAMES dataset therefore providing near-contiguous coverage of two
complete orbits. Taking this initial estimate of the orbital period as
a starting point, an iterative fit using the Levenberg-Marquardt
algorithm was used to refine the period and determine the
semi-amplitude and systemic velocity of the secondary; a
  non-zero eccentricity was allowed by the model, although all
  best fits were consistent with a circular orbit. The
resulting solution is generally an excellent fit to the spectroscopic
results (see the top panel of Fig.~\ref{fig:phase_lightcurve}), with a
period $4.20381(12)$ days, superior conjunction at
MJD~52924.21553(66), semi-amplitude $K_2=250.1\pm2.8$\kms, and a
systemic velocity $\gamma_2=160.9\pm1.9$\kms that is very close to the
local gas velocity of $162.2\pm1.8$\kms found from the strong nebular
H$\alpha$ line. Fixing the period and setting the phase as 180 degrees
away from the secondary then yields a semi-amplitude and systemic
velocity for the primary of $K_1=156.0\pm4.3$\kms and
$\gamma_1=150.8\pm3.3$\kms respectively. We therefore find a mass
ratio $q=M_2/M_1=0.62\pm0.02$,
\begin{equation}
M_1\text{sin}^3i = \frac{(1+q)^2 PK_2^3}{2\pi G} = 18.0\pm0.8 M_\odot
\end{equation}
and
\begin{equation}
M_2\text{sin}^3i = 11.2\pm0.5 M_\odot
\end{equation}

\subsection{Photometry}
\label{sec:photometry}

The \textit{nightfall}
code\footnote{http://www.hs.uni-hamburg.de/DE/Ins/Per/Wichmann/Nightfall}
was used to model the light curve of \object{NGC346-013}. Following
the analysis of Sect.~\ref{atmos}, the temperature of the secondary
was set at 24\,500K and a model atmosphere with $\log g=3.5$ was used,
while the primary was set at 34\,500K and $\log g=4$, and the radius
was set so that the B-type secondary was $1.8\times$ more luminous
than the O-star primary. Both temperatures were allowed to vary within
the error range of the {\sc tlusty} model. The eccentricity was
  initially set to zero, as implied by both the RV solution and the
  near-contact light curve, while the orbital period $P$ and mass
ratio $q$ were fixed at the values derived in Sect.~\ref{sec:radial}
and a `third light' contribution of 34\% was set as discussed in
Sect.~\ref{sec:photo}. Detailed reflection was included in the model,
and a linear limb-darkening law was assumed. Errors are internal to
the model and do not include uncertainty in the `third light'
contribution, although these do not strongly affect the derived
inclination.

The model converges to an inclination of $78.3\pm1.5$ degrees, with a
$11.2R_\odot$ secondary that has filled its Roche lobe and a
$4.6R_\odot$ primary which is almost completely eclipsed at phase 0;
we conservatively estimate errors on the radii of $\pm0.6R_\odot$
and $\pm0.5R\odot$ respectively. In such a configuration the primary
eclipse is therefore deeper than secondary eclipse, despite the
smaller primary being intrinsically less luminous. The lightcurve fit
favours a slightly lower temperature ratio than the {\sc tlusty}
model, with a 33\,500K primary and 25\,000K secondary, but poor
photometric sampling of the primary eclipse leads to some uncertainty
as to the primary temperature, and with errors of $\pm$2000K and
$\pm$1000K respectively these values are consistent with the
uncertainties of the fits to the FLAMES spectra. However, it was not
possible to reproduce the clearly-asymmetric secondary eclipse well
with the initial model, which also favoured a small degree of
eccentricity ($e\sim0.02\pm0.01$) to correctly reproduce the
eclipse timing even though a non-circular orbit is inconsistent with
the putative evolutionary state. The uneven secondary eclipse implies
a non-uniform surface brightness, and this was modelled by including a
`hotspot' on the trailing side (i.e. such that the luminosity of the
secondary is systematically higher prior to eclipse entry, when the
hotspot is visible, than after eclipse when the hotspot is not in the
line of sight). When this effect was included, the resultant fit to
the whole lightcurve was significantly improved, with a non-zero
eccentricity no longer required to reproduce the eclipse timing.  The
best-fit hotspot has a temperature ratio of $\sim1.5$ ($T\sim38$kK)
and a longitude $-30^\circ$ that imply that the hotspot is not caused
by irradiation by the hotter primary, and instead reflects a region
where the primary wind interacts with the photosphere of the
secondary\footnote{This is very similar to the effect found in the
  6.6-day contact binary \object{Cyg~OB2\#5} by \cite{linder}.}.  Fits
to the light curve both with and without the inclusion of a hotspot
are shown in the bottom panel of Fig.~\ref{fig:phase_lightcurve}, with
parameters listed in Table~\ref{tab:results}. Considerable
  scatter is present in the photometry, with some data points between
  phases $\sim$0.75 and $\sim$0.95 appearing to favour the
  `no-hotspot' model. These data were obtained over a two-year
  baseline, and we therefore speculate that variability may be
present in this feature, while the true interaction is undoubtedly
more complex than the simple hotspot considered here. However, we note
that omission of the hotspot does not significantly affect either the
derived inclination or the conclusion that the secondary has filled
its Roche lobe.

\subsection{Orbital Parameters}

\begin{table}
\caption{Summary of orbital and physical parameters of \object{NGC346-013} 
from the analysis in Sections~\ref{sec:radial} and~\ref{sec:photometry}.}
\label{tab:results}
\begin{center}
\begin{tabular}{lll}
Parameter & Value\\
\hline\hline
$T_0$ (MJD)                  & $52924.21553(66)$\\
$P$ (days)                   & $4.20381(12)$\\
$q (=M_2/M_1)$               & $0.624\pm0.019$\\
$a$ ($R_\odot$)               & $34.4\pm0.5$\\
$i$                          & $78.3^\circ\pm1.5$\\
$e$                          & $<0.02$\\
\\
                                     & Primary      & Secondary\\ 
$T_{\text{\sc tlusty}}$ (K)            & $34\,500\pm3000$ & $24\,500\pm1500$  \\
$T_{\text{\sc nightfall}}$ (K)        & $33\,500\pm2000$ & $25\,000\pm1000$  \\
$\log g$ ({\sc tlusty})          & $3.9\pm0.3$     & $3.45\pm0.2$\\
Filling factor                  &  --             & $0.98\pm0.02$\\
$R$ ($R_\odot$)                  & $4.6\pm0.5$ & $11.2\pm0.6$ \\
$\gamma$ (\kms)                 & $150.8\pm3.3$ & $160.9\pm1.9$ \\
$K$ (\kms)                      & $156.0\pm4.3$ & $250.1\pm2.8$ \\
$M$sin$^3i$ ($M_\odot$)          & $18.0\pm0.8$ & $11.2\pm0.5$ \\ 
$M$ ($M_\odot$)                  & $19.1\pm1.0$  & $11.9\pm0.6$\\
\hline
\end{tabular}
\end{center}
\end{table}

Combining the radial velocity and photometric models, we find masses
for the two components of $19.1\pm1.0M_\odot$ and $11.9\pm0.6M_\odot$
respectively, with the latter value consistent with the $\sim$B1
spectral type of the secondary, e.g. the evolutionary masses from
\cite{ih06} and \cite{t07}.  From the orbital parameters we find a
separation of $34.4R_\odot$, with a circular orbit ($e<0.02$) as
expected for a Roche-lobe filling system.  Synchronous rotation would
imply v~sin$i$ of 132\kms, somewhat faster than the v~sin$i$ of
110\kms found from the spectral line widths of the secondary; we
return to this issue in Sect.~\ref{sec:evol}. Taking the photometric
radius of $11.2\pm0.6R_\odot$ and secondary mass of
$11.9\pm0.6M_\odot$ gives $\log g=3.43$, in excellent agreement with
the spectroscopic value of $\log g=3.45$ determined in
Sect.~\ref{sec:tlusty}. The primary radius $4.6\pm0.5R_\odot$ implies
$\log g\sim 4.4\pm0.1$, somewhat higher than inferred from the model
fit to the hydrogen lines\footnote{This discrepancy may
    reflect the effects of rapid rotation and a lower effective
    equatorial surface gravity on the hydrogen line profiles.}, while
the derived radius appears rather low compared to galactic O9 dwarfs
(cf. the list of detached O binaries in \citealt{gies}, or
\object{V3903~Sag.}, a detached, short-period O7+O9 binary in which
the secondary has a near-identical mass and temperature to the primary
in \object{NGC346-013} but a radius $\sim$25\% higher;
\citealt{vaz}). However, the derived values are broadly consistent
with a $20M_\odot$ model at SMC metallicity with a zero-age main
sequence radius of $5.2R_\odot$ at $T_\text{eff}=36\,000K$ and
$\log g=4.3$ \citep{brott}, and the photometric and model atmosphere
values are in acceptable agreement given the limitations of the
photometric dataset and the paucity of spectral features from the
O-star.

\subsection{Alternative lightcurve model}

If the radius of the primary is not constrained to ensure the
brightness ratio implied by the model fits to the hydrogen lines, we
find an alternative lightcurve fit at somewhat lower inclination
($71^\circ\pm2$) in which the general properties of the B-type
secondary are broadly unchanged but the O-type primary has a radius
$12\pm1R_\odot$ and Roche-lobe filling factor $\sim0.8\pm0.1$ that
makes it slightly larger than its companion and consequently
significantly \textit{more} luminous. Revised masses for the system
then become $21.3\pm1.3M_\odot$ and $13.3\pm0.8M_\odot$ respectively.
In this configuration neither component is deeply eclipsed, while
inclusion of `hotspots' on the B-type star is again required to
recreate the profile of the secondary eclipse. However, {\sc tlusty}
model fits to the Balmer series lines discussed in Sect.~\ref{atmos}
are strongly inconsistent with a more luminous primary, and it was not
possible to achieve an acceptable fit to the FLAMES spectra with this
alternative model. Consequently we do not favour this solution, but
additional spectroscopic and photometric observations will be required
to conclusively resolve this possible discrepancy.

\section{Discussion}
\label{sec:discuss}

\subsection{Evolutionary state}
\label{sec:evol}

\begin{figure}
\begin{center}
\resizebox{\hsize}{!}{\includegraphics{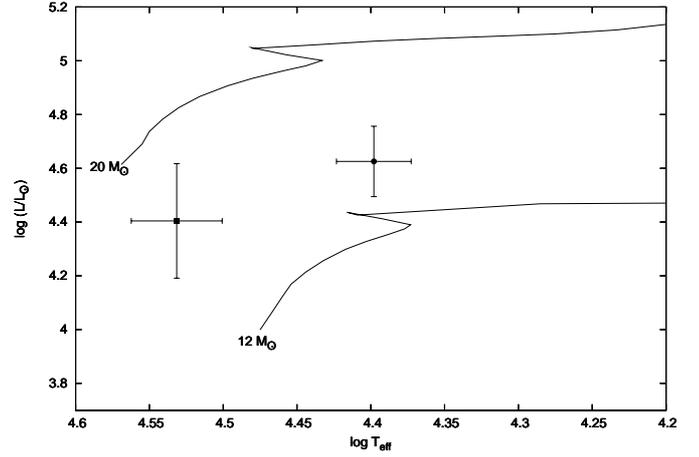}}
\caption{Location of the primary and secondary relative to illustrative 
SMC evolutionary tracks from \cite{mm01} appropriate for the \textit{current} 
masses of the two components.}
\label{fig:hrd}
\end{center}
\end{figure}

Few clues to the nature of the primary are found in our spectra, with
the majority of features within the FLAMES coverage originating in the
$\sim$B1 secondary. Modelling suggests a temperature of $\sim$34.5kK
and an $\sim$O9--9.5~V spectral type for the primary, while the high
rotational velocity of $\sim$320\kms derived from the broad
\ion{He}{ii} lines (see Sect.~\ref{sec:rotvel}) implies that it has been
`spun up' by mass transferred from the secondary, as tidal effects in
close binary systems are otherwise expected to lead to synchronization
of rotation. The secondary appears to be rotating
  somewhat slower than expected for a synchronized system, consistent
  with a predicted spin-down of the mass donor during mass transfer
  \citep{langer98, langer03}, and has almost filled its Roche lobe,
implying that despite its \textit{current} lower mass it is in a more
advanced evolutionary state. The primary appears somewhat
underluminous for its mass (see Fig.~\ref{fig:hrd}),
  consistent with a mass-gainer that has not rejuvenated completely
  \citep{braun}; significant helium enrichment indicative of a more
  evolved state would place it close to the Wolf-Rayet phase, with an
  implied luminosity of log($L$/$L_\odot$)$>$5.5 inconsistent with our
  observations. We note that the system does not feature in the X-ray
  catalogue from {\it Chandra} observations of the NGC\,346 region by
  \citet{n03}, but the expected X-ray luminosities of
  $\lsim$10$^{32}$~erg~s$^{-1}$ characteristic of individual OB stars
  \citep{clark08} are too low for detection\footnote{The faintest
    sources in \cite{n03} have fluxes of
    $\sim$6$\times10^{32}$erg~s$^{-1}$, assuming a distance of 59 kpc
    for the SMC.}  unless X-ray emission is significantly enhanced via
  wind interaction.

At first sight two stars of near-equal initial mass,
i.e. $M_{\text{ini}}\sim16+15M_\odot$, and an orbital period of a
few days would appear a simple model of the precursor to the current
system. If the initial period is very short ($P\lsim2.5$ days) such a
configuration may reach contact and merge while both stars are still
on the main sequence \citep{wellstein01, p05}, but for slightly longer
periods the system will evolve via quasi-conservative case~A mass
transfer as the initially more massive star overflows its Roche lobe,
potentially allowing the exchange of the $\sim4M_\odot$ needed to
reach the current $19+12M_\odot$ mass ratio while `spinning up' the
mass gainer to rapid rotation. However, in such a scenario mass
transfer would not halt at this point, but would continue at
very high rates ($\gsim10^{-4}M_\odot$yr$^{-1}$;
\citealt{wellstein01}) until most of the hydrogen envelope
of the mass donor is lost: the eclipsing binary \object{RY~Scuti}
appears to be possibly the only known example of this process in
action, with ongoing mass transfer from an $8M_\odot$ primary to
a $30M_\odot$ companion embedded in a dense accretion disk
\citep{smith11}. The current mass ratio of \object{NGC346-013} would
therefore place it in the middle of fast case~A transfer, but the
expected duration of this phase is so short ($\sim10^4$~years) that
direct observation is unlikely, while the extensive circumstellar
material (and possibly B[e] star morphology) indicative of rapid mass
transfer is not observed.

At the end of the rapid phase of case~A transfer, `slow' transfer
commences, lasting for $\sim1$Myr at rates
$\sim10^{-7}M_\odot$yr$^{-1}$ \citep{wellstein01,langer03}. An
initial $16+15M_\odot$ system would start the slow phase with an
overluminous low-mass donor and a rapidly-rotating accretor with
$M\sim25M_\odot$ \citep{wellstein01,langer03}, but a higher mass
initial configuration provides a much better match: taking model~13 of
\cite{wellstein} as a template, a $22+18M_\odot$ system in an inital
three-day orbit will exit the fast phase of mass transfer with an
11.5$M_\odot$ mass donor that still fills its Roche lobe, very similar
to the B-type star in \object{NGC346-013}. The model is not perfect,
as at the end of the fast phase of mass transfer both the orbital
period of 7.3 days and the $28.5M_\odot$ mass-gainer are again
inconsistent with the current state of \object{NGC346-013}, but the
model was computed assuming fully conservative mass transfer with no
loss of mass or angular momentum from the system. However, models
including rotation indicate that mass transfer may be significantly
{\it non}-conservative, with only a fraction of the transferred
material retained by the accretor while the remainder is ejected from
the system \citep{p05}. Assuming an efficiency $\beta\sim50$\% and a
slightly greater initial mass ratio of $\sim22+15M_\odot$, at the
end of the fast phase system we would again observe an $11.5M_\odot$
mass donor, but the accretor would have only increased in mass to
$\sim20M_\odot$, in good agreement with our observations, while the
loss of angular momentum from the system might be sufficient to
restrict the increase in orbital period. The B-type star would be
expected to have nitrogen enhanced by up to 1\,dex as
  CNO-processed material is exposed by the loss of the hydrogen
  envelope \citep{p05,langer08}, consistent with the high (but not
exceptional) nitrogen abundance found in Sect.~\ref{sec:abund}.

Such a scenario requires confirmation from binary evolutionary models
tailored to match the parameters of \object{NGC346-013}, but
represents an encouraging match to the observations of the system.
Tidal effects are expected to slow the rotation of the mass gainer
within a few $10^5$~years \citep{p05}, with the still-rapid rotation
of the primary therefore suggesting that fast transfer has ended
relatively recently. Within $\sim10^6$ years the system will return
to synchronous rotation, while slow case~A transfer will finish when
core hydrogen burning ends in the mass donor. A second phase of rapid
case~AB transfer then commences with the onset of shell burning,
spinning up the O-type star, still on the main sequence, back to
near-critical rotation \citep{langer03}. During this phase the mass
donor will lose its remaining hydrogen envelope to leave a helium
core, while the orbital period expands greatly \citep{wellstein01},
leaving a rapidly rotating $\sim20-25M_\odot$ OB (super)giant that
appears as an apparently-single star, with the extreme mass ratio and
increasingly low luminosity of the mass donor making detection
challenging. The initially more massive star will ultimately end its
life as a type~Ib/c supernova, forming a neutron star due to the
reduced core mass resulting from loss of the hydrogen envelope
\citep{p05} and the early termination of shell hydrogen burning
\citep{brown}. In contrast, the spun-up mass gainer may then evolve
under chemically homogeneous evolution, returning to critical
rotation at the end of core hydrogen burning and ultimately forming a
rapidly-rotating Wolf-Rayet star that may represent a future
$\gamma$-ray burst progenitor (cf. \citealt{cantiello}).

\subsection{NGC346-013 in the context of the SMC binary population}

The advent of large multi-object spectrographs has allowed the
efficient follow-up of eclipsing binaries identified in large
photometric surveys (e.g. OGLE-2; \citealt{udalski}), and accurate
parameters are now available for significant numbers of SMC binary
systems (e.g. \citealt{harries}, \citealt{hilditch}, and
\citealt{north}), while modelling of the \cite{hilditch} dataset has
been carried out by \cite{demink07}. In the context of these samples
the $19.1\pm1.0$ and $11.9\pm0.6M_\odot$ components of
\object{NGC346-013} are unexceptional, with several more massive
systems listed by \cite{hilditch}, who also identify some 28 systems
as being in a post-Roche-lobe overflow state consistent with the
quasi-conservative models of \cite{wellstein01}. However, in these
samples objects near $\sim20M_\odot$ have luminosities 0.1--0.5\,dex
higher than the O-type primary of \object{NGC346-013}, while the
320\kms projected rotational velocity is also substantially higher
than any object listed by \cite{north}, supporting the conclusion that
we are observing \object{NGC346-013} in an unusual evolutionary state.

\object{OGLE~5202153} ($P=4.61$d, $20+13M_\odot$; \citealt{harries})
appears very similar to \object{NGC346-013}, with an O9.5 primary
displaying He~II absorption that is absent in the spectrum of the
Roche-lobe filling B0.5~III secondary. The light curve is also
asymmetric, although in this case the luminosities of the two
components are similar, a result of a larger primary radius than we
infer here. \cite{demink07} model the evolution of this system, and
find non-conservative models ($\beta=0.5$) give a better fit than
fully-conservative mass transfer. Additional parameters such as
chemical abundances and the projected rotational velocity of the
primary that would confirm the similarity of \object{NGC346-013} and
\object{OGLE~5202153} are not known. Two further systems,
\object{OGLE~5300549} (P$_\text{orb}=1.33$d, $25+17M_\odot$;
\citealt{hilditch}) and \object{OGLE~9163232} (=\object{Hodge~53-47},
P$_\text{orb}=2.21$d, $26+16M_\odot$; \citealt{harries};
\citealt{morrell}) are identified by \cite{hilditch} as objects that
may have recently exited the fast phase of mass
transfer. \cite{morrell} suggest that the O6~V and O4--5~III(f)
components of \object{OGLE~9163232} have undergone significant mass
transfer, although the current 2.2-day orbital period would require a
particularly compact initial configuration, and the system may
ultimately merge as the O6~V mass gainer ends core hydrogen burning,
fills its Roche lobe, and begins reverse mass transfer while the mass
donor is still on the main sequence (\citealt{wellstein01}; see also
\citealt{demink07}). \object{OGLE~5300549} has a B0 primary that, like
\object{NGC346-013}, also seems somewhat underluminous for its mass,
while the system has the largest temperature ratio in the sample of
\cite{hilditch}. In this case the orbital period is so short that the
rotational mixing pathway of \cite{demink} is important, and
identification as a post mass-transfer system may be uncertain.

\subsection{Evolutionary implications for NGC\,346}

The systemic velocity of NGC346-013 ($\sim$155 \kms) is comparable
with the radial velocities of non-binary members of NGC\,346
\citep[e.g.][]{evans06}, while the location 2\arcmin.34 (equivalent
to 41\,pc at the distance of the SMC) from the O4~III(f)
star \object{MPG\,435}\footnote{$\alpha=$~00:59:04.49,
  $\delta=-$72:10:24.68 (J2000)} in the central core \citep{mpg}
places it well within the $\sim$7$'$ ionized region of NGC\,346
\citep{rpb}\footnote{The radial distance from the centre of the
  cluster was given by \citet{evans06} as 1.89$'$; this distance
  was calculated taking coordinates for the centre of the cluster from
  {\sc simbad} (http://simbad.u-strasbg.fr).}. \object{NGC346-013}
therefore appears to be a true member of the cluster and does not
appear to have been ejected.

\cite{sabbi} report extensive subclustering in \object{NGC\,346}, with
\object{NGC346-013} apparently associated with SC-16, a subcluster
with radius 1.6pc that likely pre-dates the formation of the main
\object{NGC\,346} cluster. An age of $15\pm2.5$Myr is reported for
SC-16 by \cite{sabbi}, consistent with independent determinations of
5-15Myr \citep{h08} and 12.5-18Myr \citep{cig10a}, but notably
discrepant with ages of $3\pm1$Myr for the subclusters that trace the
major epoch of star formation in \object{NGC\,346}. However, ages of
$\ge$10Myr are incompatible with \object{NGC346-013} being a {\it bona
  fide} member of SC-16, as even with minimum progenitor masses the
system would start to interact at an intermediate age ($\sim8$Myr;
\citealt{langer08}) and comparison with SMC evolutionary tracks
\citep{mm01} shows that core hydrogen burning would have ended at
$\sim10$Myr, while at the ages reported by \cite{sabbi} and
\cite{cig10a} core helium burning would also have ended\footnote{While
  the inclusion of rotation in the evolutionary tracks results in core
  hydrogen burning ending slightly later, the expectation that a
  short-period binary will reach synchronous rotation implies that the
  tracks with rapid rotation do not apply.}. Higher progenitor masses
would imply earlier interaction, and it therefore seems possible that
\object{NGC346-013} represents a chance association with SC-16, and is
more likely to have an age compatible with the main burst of star
formation in \object{NGC\,346}. However, a young age for SC-16 might
nevertheless be consistent with \object{NGC346-013} being a member,
and any conclusive determination of the likely age and formation
scenario for \object{NGC346-013} and its consequent implications for
the star formation history of \object{NGC\,346} must also await
tailored modelling with binary evolution codes.

\section{Conclusions and future work}
\label{sec:conc}

We find current masses of $19.1\pm1.0M_\odot$ and
$11.9\pm0.6M_\odot$ for the O- and B-type components of
\object{NGC346-013} respectively, with comparison with tailored
non-LTE model atmosphere spectra favouring a more luminous B-type star
and consequently a compact O-type primary of
$4.6\pm0.5R_\odot$. Simple conservative mass transfer in an initial
16+15$M_\odot$ system appears unlikely, and we suggest that the system
began life as a $\sim$22+15$M_\odot$ binary, with the current B-type
secondary initially the more massive star, and has recently exited a
phase of rapid, non-conservative mass transfer that has spun the
current primary up to high rotation. This scenario suggests an age of
$\lsim$6Myr consistent with the recent burst of star formation within
\object{NGC\,346}. Within $\sim$1--2$\times$10$^6$ years the system
will undergo a second phase of fast case~AB mass transfer to form a
low-mass helium star + OB (super)giant binary with a period of months.
The initially more massive star will ultimately form a neutron star
and, if the system is not disrupted by a supernova `kick', a
long-period neutron star + OB supergiant high-mass X-ray binary, while
the mass gainer may subsequently evolve at high rotation and may be a
possible $\gamma$-ray burst progenitor.

Further optical observations will allow a search for evidence of other
spectral features originating from the primary, in order to better
constrain its nature; in particular ultra-violet data could offer
insight, as the O-type star would be expected to dominate the flux in
that region. High quality photometric data would also resolve the
uncertainty as to the radius of the primary, and would test the
robustness of the orbital parameters presented here. However, despite
this uncertainty the general properties of the system are well
constrained, and it therefore represents an excellent candidate for
tailored modelling with binary evolution codes.

\begin{acknowledgements}
We thank William Taylor for reducing the 2008 FLAMES dataset, Philip
Dufton and Robert Ryans for use of the Queens {\sc tlusty} grid, and
Myron Smith for valuable comments. We also thank the referee Pierre
North for a detailed and helpful report, and the editor Ralf
Napiwotzki for suggesting clarifications to the manuscript. This
research has made use of the SIMBAD database, operated at CDS,
Strasbourg, France.
\end{acknowledgements}

\end{document}